\documentclass[10pt,twocolumn]{article}
\usepackage[top=1.8cm, bottom=1.8cm, left=1.8cm, right=1.8cm]{geometry}

\usepackage{amsmath,amsthm,amssymb,amsfonts}
\usepackage{graphicx}
\usepackage{times}
\usepackage{algorithm}

\usepackage{multirow}

\usepackage{color}
\definecolor{darkgreen}{RGB}{0,90,25}
\definecolor{darkblue}{RGB}{0,0,75}
\usepackage[bookmarks=false, colorlinks=true, plainpages = false,  linkcolor = darkgreen,   citecolor = darkgreen, urlcolor = darkblue, filecolor = black]{hyperref}
\usepackage{algpseudocode}
\usepackage{cleveref}
\usepackage{flushend}
\usepackage[small,bf]{caption}
\usepackage{subcaption}
\usepackage[compact]{titlesec}
\titlespacing*{\section}{0pt}{*4}{4pt} 
\titlespacing{\subsection}{0pt}{*2}{2pt}
\titlespacing{\subsubsection}{0pt}{*2}{2pt}

\renewcommand{\footnoterule}{%
  \kern -3pt
  \hrule width 1in 
  \kern 2pt
}

\usepackage[sort,numbers]{natbib}

\makeatletter
\newcounter{algorithmicH}%
\let\oldalgorithmic\algorithmic
\renewcommand{\algorithmic}{%
  \stepcounter{algorithmicH}%
  \oldalgorithmic}%
\renewcommand{\theHALG@line}{ALG@line.\thealgorithmicH.\arabic{ALG@line}}
\makeatother

\newcommand{\paragraphbe}[1]{\vspace{0.75ex}\noindent{\bf \em #1}\hspace*{.3em}}

\algdef{SE}[SUBALG]{Indent}{EndIndent}{}{\algorithmicend\ }%
\algtext*{Indent}
\algtext*{EndIndent}

\newcommand{\bprop}{b_\text{prop}} %
\newcommand{\bnonprop}{b_\text{nonprop}} %
\newcommand{\params}{\theta} %
\newcommand{\calX}{\mathcal{X}}
\newcommand{\calY}{\mathcal{Y}}

\renewenvironment{thebibliography}[1]{
  \begin{oldthebibliography}{#1}
    \setlength{\itemsep}{0.1em}
    \setlength{\parskip}{0.1em}
}
{
  \end{oldthebibliography}
}

\usepackage[hang,flushmargin]{footmisc}

\makeatletter
\def\url@leostyle{%
  \@ifundefined{selectfont}{\def\UrlFont{}}%
  {\def\UrlFont{}}%
}
\makeatother
\usepackage[hyphenbreaks]{breakurl}

\makeatletter
\def\@copyrightspace{\relax}
\makeatother
\begin{document} 

\title{\bf Exploiting Unintended Feature Leakage in Collaborative Learning$^*$}

\date{} 

\author{\fontsize{11}{13}\selectfont Luca Melis$^\dag$\\
\fontsize{11}{13}\selectfont UCL\\
\hspace*{-0.5cm} \fontsize{11}{13}\selectfont luca.melis.14@alumni.ucl.ac.uk \hspace*{-0.5cm}
\and
\fontsize{11}{13}\selectfont Congzheng Song$^\dag$\\
\fontsize{11}{13}\selectfont Cornell University\\
\hspace*{-0.5cm} \fontsize{11}{13}\selectfont cs2296@cornell.edu \hspace*{-0.5cm} 
\and
\fontsize{11}{13}\selectfont Emiliano De Cristofaro\\
\fontsize{11}{13}\selectfont UCL \& Alan Turing Institute\\
\hspace*{-0.5cm} \fontsize{11}{13}\selectfont e.decristofaro@ucl.ac.uk \hspace*{-0.5cm} 
\and
\fontsize{11}{13}\selectfont Vitaly Shmatikov\\
\fontsize{11}{13}\selectfont Cornell Tech\\
\hspace*{-0.5cm} \fontsize{11}{13}\selectfont shmat@cs.cornell.edu \hspace*{-0.5cm} 
}

\maketitle

\renewcommand*{\thefootnote}{\fnsymbol{footnote}}
\footnotetext{$^*$In Proceedings of 40th IEEE Symposium on Security \& Privacy (S\&P 2019).\\
$^\dag$Authors contributed equally.}
\renewcommand*{\thefootnote}{\arabic{footnote}}

\maketitle

\begin{abstract}
Collaborative machine learning and related techniques such as federated
learning allow multiple participants, each with his own training dataset,
to build a joint model by training locally and periodically exchanging
model updates.
We demonstrate that these updates leak \emph{unintended} information about
participants' training data and develop passive and active inference
attacks to exploit this leakage.  First, we show that an adversarial
participant can infer the presence of exact data points\textemdash for
example, specific locations\textemdash in others' training data (i.e.,
\emph{membership} inference).  Then, we show how this adversary can infer
\emph{properties} that hold only for a subset of the training data and
are independent of the properties that the joint model aims to capture.
For example, he can infer when a specific person first appears in
the photos used to train a binary gender classifier.
We evaluate our attacks on a variety of tasks, datasets, and learning
configurations, analyze their limitations, and discuss possible defenses.
\end{abstract}

\section{Introduction}

Collaborative machine learning (ML) has recently emerged as an
alternative to conventional ML methodologies where all training data is
pooled and the model is trained on this joint pool.  It allows two or
more participants, each with his own training dataset, to construct a
joint model.  Each participant trains a local model on his own data and
periodically exchanges model parameters, updates to these parameters,
or partially constructed models with the other participants.

Several architectures have been proposed for distributed,
collaborative, and federated learning~\cite{dean2012large,
chilimbi2014project, xing2015petuum, moritz2015sparknet, lin2017deep,
zinkevich2010parallelized}: with and without a central server,
with different ways of aggregating models, etc.  The main
goal is to improve the training speed and reduce overheads,
but protecting privacy of the participants' training data is
also an important motivation for several recently proposed
techniques~\cite{mcmahan2016communication, shokri2015privacy}.
Because the training data never leave the participants' machines,
collaborative learning may be a good match for the scenarios where
this data is sensitive (e.g., health-care records, private images,
personally identifiable information, etc.).  Compelling applications
include training of predictive keyboards on character sequences
that users type on their smartphones~\cite{mcmahan2016communication},
or using data from multiple hospitals to develop predictive models
for patient survival~\cite{jochems2} and side effects of medical
treatments~\cite{jochems1}.

Collaborative training, however, {\em does} disclose information via model
updates that are based on the training data.  The key question we
investigate in this paper is: \textbf{what can be inferred about a
participant's training dataset from the model updates} revealed during
collaborative model training?

Of course, the purpose of ML is to discover new information about the
data.  Any useful ML model reveals something about the population from
which the training data was drawn.  For example, in addition to accurately
classifying its inputs, a classifier model may reveal the features that
characterize a given class or help construct data points that belong to
this class.  In this paper, we focus on inferring ``unintended'' features,
i.e., properties that hold for certain subsets of the training data,
but not generically for all class members.

The basic privacy violation in this setting is \emph{membership
inference}: given an exact data point, determine if it was used to train
the model.  Prior work described passive and active membership inference
attacks against ML models~\cite{shokri2017membership, hayes2017logan},
but collaborative learning presents interesting new avenues for such
inferences.  For example, we show that an adversarial participant can
infer whether a specific location profile was used to train a gender
classifier on the FourSquare location dataset~\cite{yang2015participatory}
with 0.99 precision and perfect recall.

We then investigate passive and active \emph{property inference} attacks
that allow an adversarial participant in collaborative learning to infer
properties of other participants' training data that are not true of
the class as a whole, or even \emph{independent} of the features that
characterize the classes of the joint model.  We also study variations
such as inferring when a property appears and disappears in the data
during training\textemdash for example, identifying when a certain person
first appears in the photos used to train a generic gender classifier.

For a variety of datasets and ML tasks, we demonstrate successful
inference attacks against two-party and multi-party collaborative
learning based on~\cite{shokri2015privacy} and multi-party federated
learning based on~\cite{mcmahan2016communication}.  For example,
when the model is trained on the LFW dataset~\cite{huang2007labeled}
to recognize gender or race, we infer whether people in the training
photos wear glasses\textemdash a property that is {\em uncorrelated}
with the main task.  By contrast, prior property inference
attacks~\cite{ateniese2015hacking, hitaj2017deep} infer only properties
that characterize an entire class.  We discuss this critical distinction
in detail in Section~\ref{sec:privML}.

Our key observation, concretely illustrated by our experiments, is that
modern deep-learning models come up with separate internal representations
of all kinds of features, some of which are independent of the task
being learned.  These ``unintended'' features leak information about
participants' training data.  We also demonstrate that an \emph{active}
adversary can use multi-task learning to trick the joint model into
learning a better internal separation of the features that are of interest
to him and thus extract even more information.

Some of our inference attacks have direct privacy implications.
For example, when training a binary gender classifier on the
FaceScrub~\cite{ng2014data} dataset, we infer with high accuracy (0.9
AUC score) that a certain person appears in a single training batch even
if half of the photos in the batch depict other people.  When training a
generic sentiment analysis model on Yelp healthcare-related reviews, we
infer the specialty of the doctor being reviewed with perfect accuracy.
On another set of Yelp reviews, we identify the author even if their
reviews account for less than a third of the batch.

We also measure the performance of our attacks vis-\`a-vis the
number of participants (see Section~\ref{sec:experimentsM}).  On the
image-classification tasks, AUC degrades once the number of participants
exceeds a dozen or so.  On sentiment-analysis tasks with Yelp reviews,
AUC of author identification remains high for many authors even with
30 participants.

Federated learning with model averaging~\cite{mcmahan2016communication}
does not reveal individual gradient updates, greatly reducing the
information available to the adversary.  We demonstrate successful
attacks even in this setting, e.g., inferring that photos of a certain
person appear in the training data. %

Finally, we evaluate possible defenses\textemdash sharing fewer gradients,
reducing the dimensionality of the input space, dropout\textemdash
and find that they do not effectively thwart our attacks.  We also attempt to use
participant-level different privacy~\cite{mcmahan2017learning}, which, however,
is geared to work with thousands of users, and the joint model fails
to converge in our setting.

\section{Background}\label{sec:background}

\subsection{Machine learning (ML)}

An ML model is a function $f_\params:\calX\mapsto\calY$ parameterized by
a set of \emph{parameters} $\params$, where $\calX$ denotes the input
(or feature) space, and $\calY$ the output space.  

In this paper, we focus on the supervised learning of classification
tasks.  The \emph{training data} is a set of data points labeled
with their correct classes.  We work with models that take as input
images or text (i.e., sequences of words) and output a class label.
To find the optimal set of parameters that fits the training data,
the training algorithm optimizes the \emph{objective (loss) function},
which penalizes the model when it outputs a wrong label on a data point.
We use $L(x, y;\theta)$ to denote the loss computed on a data point $(x,
y)$ given the model parameters $\theta$, and $L(b;\theta)$ to denote
the average loss computed on a batch $b$ of data points.

\paragraphbe{Stochastic Gradient Descent (SGD).} 
There are many methods to optimize the objective function.  Stochastic
gradient descent (SGD) and its variants are commonly used to train
artificial neural networks, but our inference methodology is not
specific to SGD.  SGD is an iterative method where at each step the
optimizer receives a small batch of the training data and updates the
model parameters $\theta$ according to the direction of the negative
gradient of the objective function with respect to $\theta$ and scaled
by the learning rate $\eta$.  Training finishes when the model has
converged to a local minimum, where the gradient is close to zero.
The trained model is tested using held-out data, which was not used
during training.  A standard metric is \emph{test accuracy}, i.e.,
the percentage of held-out data points that are classified correctly.

\paragraphbe{Hyperparameters.} 
Most modern ML algorithms have a set of tunable hyperparameters,
distinct from the model parameters.  They control the number of training
iterations, the ratio of the regularization term in the loss function
(its purpose is to prevent overfitting, i.e., a modeling error that occurs
when a function is too closely fitted to a limited set of data points),
the size of the training batches, etc.

\paragraphbe{Deep learning (DL).} 
A family of ML models known as deep learning recently became very
popular for many ML tasks, especially related to computer vision and
image recognition~\cite{schmidhuber2015deep,lecun2015deep}.  DL models
are made of layers of non-linear mappings from input to intermediate
hidden states and then to output.  Each connection between layers has a
floating-point weight matrix as parameters.  These weights are updated
during training.  The topology of the connections between layers is
task-dependent and important for the accuracy of the model.

\subsection{Collaborative learning}
\label{sec:collabml}

Training a deep neural network on a large dataset can be
time- and resource-consuming.  A common scaling approach is
to partition the training dataset, concurrently train separate
models on each subset, and exchange parameters via a parameter
server~\cite{chilimbi2014project,dean2012large}.  During training,
each local model pulls the parameters from this server, calculates the
updates based on its current batch of training data, then pushes these
updates back to the server, which updates the global parameters.

Collaborative learning may also involve participants who
want to hide their training data from each other.  We review
two architectures for privacy-preserving collaborative
learning based on, respectively,~\cite{shokri2015privacy}
and~\cite{mcmahan2016communication}.

\begin{algorithm}[t]
\caption{Parameter server with synchronized SGD}
\small
\begin{algorithmic}
\State \textbf{Server executes:} 
\Indent
\State Initialize $\params_0$
\For{$t=1$ to $T$}
\For{each client $k$}
\State $g_t^k\gets$\textbf{ClientUpdate}($\params_{t-1}$)
\EndFor
\State $\params_t\gets \params_{t-1} - \eta \sum_k g_t^k$ \Comment{synchronized gradient updates}
\EndFor
\EndIndent
\State
\State \textbf{ClientUpdate}($\params$): 
\Indent
\State Select batch $b$ from client's data
\State \textbf{return} local gradients $\nabla L(b;\params)$
\EndIndent
\end{algorithmic}
\label{alg:ps}
\end{algorithm}

\paragraphbe{Collaborative learning with synchronized gradient updates.}
Algorithm~\ref{alg:ps} shows collaborative learning with synchronized
gradient updates~\cite{shokri2015privacy}.  In every iteration
of training, each participant downloads the global model from the
parameter server, locally computes gradient updates based on one batch
of his training data, and sends the updates to the server.  The server
waits for the gradient updates from all participants and then applies
the aggregated updates to the global model using stochastic gradient
descent (SGD).  

In~\cite{shokri2015privacy}, each client may share only a fraction
of his gradients.  We evaluate if this mitigates our attacks in
Section~\ref{sec:grad_frac}.  Furthermore, \cite{shokri2015privacy}
suggests differential privacy to protect gradient updates.  We do
not include differential privacy in our experiments.  By definition,
record-level differential privacy bounds the success of membership
inference, but does not prevent property inference that applies to a
group of training records.  Participant-level differential privacy,
on the other hand, bounds the success of all attacks considered in
this paper, but we are not aware of any participant-level differential
privacy mechanism that enables collaborative learning of an accurate
model with a small number of participants.  We discuss this further in
Section~\ref{ssec:dp_defense}.

\begin{algorithm}[t]
\caption{Federated learning with model averaging}
\small
\begin{algorithmic}
\State \textbf{Server executes:} 
\Indent
\State Initialize $\params_0$
\State $m \gets max(C \cdot K, 1)$
\For{$t=1$ to $T$}
\State $S_t \gets \text{(random set of m clients)}$
\For{each client $k \in S_t$}
\State $\params_t^k\gets$\textbf{ClientUpdate}($\params_{t-1}$)
\EndFor
\State $\params_t\gets\sum_{k} \frac{n^k}{n} \params_t^k$ \Comment{averaging local models}
\EndFor
\EndIndent
\State
\State \textbf{ClientUpdate}($\params$): 
\Indent
\For{each local iteration}
\For{each batch $b$ in client's split}
\State $\params \gets \params - \eta \nabla L(b;\params)$
\EndFor
\EndFor
\State \textbf{return} local model $\params$
\EndIndent
\end{algorithmic}
\label{alg:fl}
\end{algorithm}

\paragraphbe{Federated learning with model averaging.} 
Algorithm~\ref{alg:fl} shows the federated learning
algorithm~\cite{mcmahan2016communication}.  We set $C$, the fraction of
the participants who update the model in each round, to $1$ (i.e., the
server takes updates from all participants), to simplify our experiments
and because we ignore the efficiency of the learning protocol when
analyzing the leakage.

In each round, the $k$-th participant locally takes several steps of
SGD on the current model using his entire training dataset of size
$n^k$ (i.e., the globally visible updates are based not on batches
but on participants' entire datasets).  In Algorithm~\ref{alg:fl},
$n$ is the total size of the training data, i.e., the sum of all $n^k$.
Each participant submits the resulting model to the server, which computes
a weighted average.  The server evaluates the resulting joint model on
a held-out dataset and stops training when performance stops improving.

The convergence rate of both collaborative learning approaches heavily
depends on the learning task and the hyperparameters (e.g., number of
participants and batch size).

\section{Reasoning about Privacy in Machine\\[-0.25ex] Learning}
\label{sec:privML}

If a machine learning (ML) model is useful, it must reveal information
about the data on which it was trained~\cite{dworknaor}.  To argue that
the training process and/or the resulting model violate ``privacy,''
it is not enough to show that the adversary learns something new about
the training inputs.  At the very least, the adversary must learn
\emph{more} about the training inputs than about other members of their
respective classes.  To position our contributions in the context of
related work (surveyed in~\Cref{sec:related}) and motivate the need to
study unintended feature leakage, we discuss several types of adversarial
inference previously considered in the research literature.

\subsection{Inferring class representatives}
\label{badprior}

Given black-box access to a classifier model, \emph{model inversion}
attacks~\cite{fredrikson2015model} infer features that characterize each
class, making it possible to construct representatives of these classes.

In the special case\textemdash and only in this special case\textemdash
where all class members are similar, the results of model inversion are
similar to the training data.  For example, in a facial recognition
model where each class corresponds to a single individual, all class
members depict the same person.  Therefore, the outputs of model
inversion are visually similar to any image of that person, including
the training photos.  If the class members are \emph{not} all visually
similar, the results of model inversion do not look like the training
data~\cite{shokri2017membership}.

If the adversary actively participates in training the model (as in
the collaborative and federated learning scenarios considered in this
paper), he can use GANs~\cite{goodfellow2014generative} to construct class
representatives, as done by Hitaj et al.~\cite{hitaj2017deep}.  Only in
the special case where all class members are similar, GAN-constructed
representatives are similar to the training data.  For example, all
handwritten images of the digit `9' are visually similar.  Therefore, a
GAN-constructed image for the `9' class looks similar to \emph{any} image
of digit 9, including the training images.  In a facial recognition model,
too, all class members depict the same person.  Hence, a GAN-constructed
face looks similar to any image of that person, including the training
photos.

Note that \textit{neither technique reconstructs actual training inputs.}
In fact, there is no evidence that GANs, as used in~\cite{hitaj2017deep},
can even distinguish between a training input and a random member of
the same class.

\begin{figure}[t]
\centering
\includegraphics[width=0.11\textwidth]{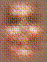}
\includegraphics[width=0.11\textwidth]{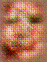}
\includegraphics[width=0.11\textwidth]{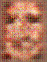}
\includegraphics[width=0.11\textwidth]{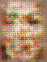}
\caption{Samples from a GAN attack on a gender classification model
where the class is ``female.''\label{ganimage}}
\vspace*{-0.4cm}
\end{figure}

Data points produced by model inversion and GANs are similar to the
training inputs only if all class members are similar, as is the case for
MNIST (the dataset of handwritten digits used in~\cite{hitaj2017deep})
and facial recognition.  This simply shows that ML works as it should.
A trained classifier reveals the input features characteristic of each
class, thus enabling the adversary to sample from the class population.
For instance, Figure~\ref{ganimage} shows GAN-constructed images for
the gender classification task on the LFW dataset, which we use in our
experiments (see~\Cref{sec:experiments2}).  These images show a generic
female face, but there is no way to tell from them whether an image of
a \emph{specific} female was used in training or not.

Finally, the active attack in~\cite{hitaj2017deep} works by overfitting
the joint model's representation of a class to a single participant's
training data.  This assumes that the entire training corpus for a given
class belongs to that participant.  We are not aware of any deployment
scenario for collaborative learning where this is the case.  By contrast,
we focus on a more realistic scenario where the training data for each
class are distributed across multiple participants, although there may
be significant differences between their datasets.

\subsection{Inferring membership in training data}
\label{subsec:membership-def}

The (arguably) simplest privacy breach is, given a model and
an exact data point, inferring whether this point was used to train the model
or not.  Membership inference attacks against aggregate statistics are
well-known~\cite{homer2008resolving,pyrgelis2017knock,dwork2015robust},
and recent work demonstrated black-box membership inference against ML
models~\cite{shokri2017membership, demyst2018, long2018understanding,
hayes2017logan}, as discussed in \Cref{sec:related}.

The ability of an adversary to infer the presence of a specific data
point in a training dataset constitutes an immediate privacy threat if
the dataset is in itself sensitive.  For example, if a model was trained
on the records of patients with a certain disease, learning that an
individual's record was among them directly affects his or her privacy.
Membership inference can also help demonstrate inappropriate uses of data
(e.g., using health-care records to train ML models for unauthorized
purposes~\cite{bbc}), enforce individual rights such as the ``right to
be forgotten,'' and/or detect violations of data-protection regulations
such as the GDPR~\cite{GDPR}.  Collaborative learning presents interesting
new avenues for such inferences.

\subsection{Inferring properties of training data}

In collaborative and federated learning, participants' training data
may not be identically distributed.  Federated learning is explicitly
designed to take advantage of the fact that participants may have private
training data that are different from the publicly available data for
the same class~\cite{mcmahan2016communication}.

Prior work~\cite{fredrikson2015model, hitaj2017deep, ateniese2015hacking}
aimed to infer properties that characterize an entire class: for example,
given a face recognition model where one of the classes is Bob, infer
what Bob looks like (e.g., Bob wears glasses).  It is not clear that
hiding this information in a good classifier is possible or desirable.

By contrast, we aim to infer \emph{properties that are true of a subset
of the training inputs but not of the class as a whole}.  For instance, when
Bob's photos are used to train a gender classifier, we infer that Alice
appears in some of the photos.  We especially focus on the properties
that are \emph{independent} of the class's characteristic features.
In contrast to the face recognition example, where ``Bob wears glasses''
is a characteristic feature of an entire class, in our gender classifier
study we infer whether people in Bob's photos wear glasses\textemdash
even though wearing glasses has no correlation with gender.  There is
no legitimate reason for a model to leak this information; it is purely
an artifact of the learning process.

A participant's contribution to each iteration of collaborative learning
is based on a batch of his training data.  We infer \emph{single-batch
properties}, i.e., detect that the data in a given batch has the property
but other batches do not.  We also infer \emph{when a property appears in
the training data}.  This has serious privacy implications.  For instance,
we can infer when a certain person starts appearing in a participant's
photos or when the participant starts visiting a certain type of doctors.
Finally, we infer properties that characterize a participant's entire
dataset (but not the entire class), e.g., authorship of the texts used
to train a sentiment-analysis model.

\begin{figure*}[t]
\centering
\includegraphics[width=0.85\textwidth]{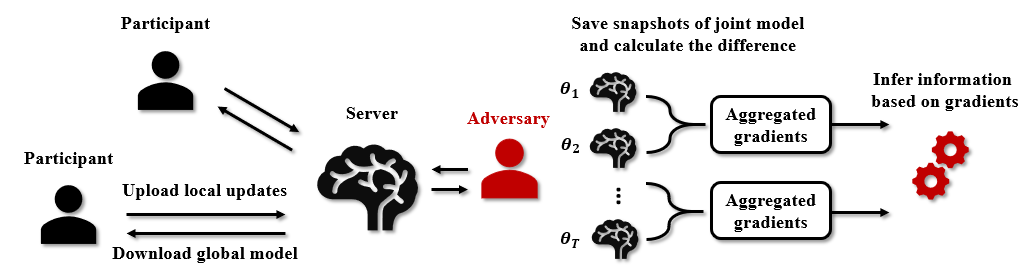}
\caption{Overview of inference attacks against collaborative learning.}
\label{fig:attack_overview}
\end{figure*}

\section{Inference Attacks}
\label{sec:inference}

\subsection{Threat model}
\label{sec:threatmodel}

We assume that $K$ participants (where $K\geq2$) jointly train
an ML model using one of the collaborative learning algorithms
described in Section~\ref{sec:collabml}.  One of the participants
is the \emph{adversary}.  His goal is to infer information about
the training data of another, \emph{target participant} by analyzing
periodic updates to the joint model during training.  Multi-party ($K>2$)
collaborative learning also involves honest participants who are neither
the adversary, nor the target.  In the multi-party case, the identities
of the participants may not be known to the adversary.  Even if the
identities are known but the models are aggregated, the adversary may
infer something about the training data but not trace it to a specific
participant; we discuss this further in Section~\ref{sec:attribution}.

The updates that adversary observes and uses for inference depend on
both $K$ and how collaborative training is done.

As inputs to his inference algorithms, the adversary uses the
model updates revealed in each round of collaborative training.
For synchronized SGD~\cite{shokri2015privacy} with $K=2$, the
adversary observes gradient updates computed on a single batch of
the target's data.  If $K>2$, he observes an aggregation of gradient
updates from all other participants (each computed on a single batch of
the respective participant's data).  For federated learning with model
averaging~\cite{mcmahan2016communication}, the observed updates are the
result of two-step aggregation: (1) every participant aggregates the
gradients computed on each local batch, and (2) the server aggregates
the updates from all participants.

For property inference, the adversary needs auxiliary training data
correctly labeled with the property he wants to infer (e.g., faces labeled
with ages if the goal is to infer ages).  For active property inference
(Section~\ref{subsec:active-att}), these auxiliary data points must also
be labeled for the main task (e.g., faces labeled with identities for
a facial recognition model).

\subsection{Overview of the attacks}

\Cref{fig:attack_overview} provides a high-level overview of our inference
attacks.  At each iteration $t$ of training, the adversary downloads the
current joint model, calculates gradient updates as prescribed by the
collaborative learning algorithm, and sends his own updates to the server.
The adversary saves the snapshot of the joint model parameters $\theta_t$.
The difference between the consecutive snapshots $\Delta\theta_t =
\theta_t - \theta_{t-1} = \sum_{k}\Delta\theta^k_t$ is equal to the
aggregated updates from all participants, hence $\Delta\theta_t -
\Delta\theta^\text{adv}_t$ are the aggregated updates from all
participants other than the adversary.

\paragraphbe{Leakage from the embedding layer.} 
\label{sec:embedleak}
All deep learning models operating on non-numeric data where the
input space is discrete and sparse (e.g., natural-language text or
locations) first use an embedding layer to transform inputs into a
lower-dimensional vector representation.  For convenience, we use
\emph{word} to denote discrete tokens, i.e., actual words or specific
locations.  Let vocabulary $V$ be the set of all words.  Each word in
the training data is mapped to a word-embedding vector via an embedding
matrix $W_\text{emb}\in\mathbb{R}^{|V|\times d}$, where $|V|$ is the
size of the vocabulary and $d$ is the dimensionality of the embedding.

During training, the embedding matrix is treated as a parameter of the
model and optimized collaboratively.  The gradient of the embedding
layer is sparse with respect to the input words: given a batch of text,
the embedding is updated only with the words that appear in the batch.
The gradients of the other words are zeros.  This difference directly
reveals which words occur in the training batches used by the honest
participants during collaborative learning.

\paragraphbe{Leakage from the gradients.} 
In deep learning models, gradients are computed by back-propagating
the loss through the entire network from the last to the first layer.
Gradients of a given layer are computed using this layer's features
and the error from the layer above.  In the case of sequential fully
connected layers $h_l, h_{l+1}$ ($h_{l+1} = W_l\cdot h_l$, where $W_l$
is the weight matrix), the gradient of error $E$ with respect to
$W_l$ is computed as $\frac{\partial E}{\partial W_l}=\frac{\partial
E}{\partial h_{l+1}} \cdot h_l$.  The gradients of $W_l$ are inner
products of the error from the layer above and the features $h_{l}$.
Similarly, for a convolutional layer, the gradients of the weights are
convolutions of the error from the layer above and the features $h_{l}$.
Observations of gradient updates can thus be used to infer feature values,
which are in turn based on the participants' private training data.

\subsection{Membership inference}
\label{subsec:membership-att}

As explained above, the non-zero gradients of the embedding layer reveal
which words appear in a batch.  This helps infer whether a given text
or location appears in the training dataset or not.  Let $V_t$ be the
words included in the updates $\Delta\theta_t$.  During training, the
attacker collects a vocabulary sequence $[V_1, \dots, V_T]$.  Given a
text record $r$, with words $V_r$, he can test if $V_r\subseteq V_t$,
for some $t$ in the vocabulary sequence.  If $r$ is in target's dataset,
then $V_r$ will be included in at least one vocabulary from the sequence.
The adversary can use this to decide whether $r$ was a member or not.

\subsection{Passive property inference}
\label{subsec:passive-att}

We assume that the adversary has auxiliary data consisting of the data
points that have the property of interest ($D_\text{prop}^\text{adv}$) and
data points that do not have the property ($D_\text{nonprop}^\text{adv}$).
These data points need to be sampled from the same class as the target
participant's data, but otherwise can be unrelated.

The intuition behind our attack is that the adversary can leverage the
snapshots of the global model to generate aggregated updates based on
the data with the property and updates based on the data without the
property.  This produces labeled examples, which enable the adversary
to train a binary \emph{batch property classifier} that determines if
the observed updates are based on the data with or without the property.
This attack is \emph{passive}, i.e., the adversary observes the updates
and performs inference without changing anything in the local or global
collaborative training procedure.

\paragraphbe{Batch property classifier.} 
Algorithm~\ref{alg:train_infer} shows how to build a batch property
classifier during collaborative training.  Given a model snapshot
$\theta_t$, calculate gradients $g_\text{prop}$ based on a batch with
the property $\bprop^\text{adv} \subset D_\text{prop}^\text{adv}$
and $g_\text{nonprop}$ based on a batch without the property
$\bnonprop^\text{adv} \subset D_\text{nonprop}^\text{adv}$.  Once enough
labeled gradients have been collected, train a binary classifier
$f_\text{prop}$.

For the property inference attacks that exploit the
embedding-layer gradients (e.g., the attack on the Yelp dataset in
Section~\ref{yelpattack}), we use a logistic regression classifier.
For all other property inference attacks, we experimented with logistic
regression, gradient boosting, and random forests.  Random forests
with $50$ trees performed the best.  The input features in this case
correspond to the observed gradient updates.  The number of the features
is thus equal to the model's parameters, which can be very large for a
realistic model.  To downsample the features representation, we apply
the max pooling operator~\cite{goodfellow2016deep} on the observed
gradient updates.  More specifically, max pooling performs a max filter
to non-overlapping subregions of the initial features representation,
thus reducing the computational cost of the attack.

{\setlength{\textfloatsep}{6pt}%
\begin{algorithm}[t]
\caption{Batch Property Classifier}
\small
\begin{algorithmic}
\State \textbf{Inputs:} Attacker's auxiliary data $D_\text{prop}^\text{adv}, D_\text{nonprop}^\text{adv}$
\State \textbf{Outputs:} Batch property classifier $f_\text{prop}$
\State $G_\text{prop}\gets\emptyset$\Comment{Positive training data for property inference}
\State $G_\text{nonprop}\gets\emptyset$\Comment{Negative training data for property inference}
\For{$i = 1$ to $T$}
\State Receive $\theta_t$ from server
\State Run \textbf{ClientUpdate}($\params_t$)
\State Sample $\bprop^\text{adv} \subset D_\text{prop}^\text{adv}, \bnonprop^\text{adv} \subset D_\text{nonprop}^\text{adv}$
\State Calculate $g_\text{prop}=\nabla L(\bprop^\text{adv}; \theta_t), g_\text{nonprop}=\nabla L(\bnonprop^\text{adv}; \theta_t)$
\State $G_\text{prop}\gets G_\text{prop}\cup \{g_\text{prop}\}$
\State $G_\text{nonprop}\gets G_\text{nonprop}\cup \{g_\text{nonprop}\}$
\EndFor
\State Label $G_\text{prop}$ as positive and $G_\text{nonprop}$ as negative
\State Train a binary classifier $f_\text{prop}$ given $G_\text{prop}, G_\text{nonprop}$
\end{algorithmic} 
\label{alg:train_infer}
\end{algorithm}}

\paragraphbe{Inference algorithm.}
As collaborative training progresses, the adversary observes gradient
updates $g_\text{obs} = \Delta\theta_t - \Delta\theta^\text{adv}_t$.
For single-batch inference, the adversary simply feeds the observed
gradient updates to the batch property classifier $f_\text{prop}$.

This attack can be extended from single-batch properties to the target's
entire training dataset.  The batch property classifier $f_\text{prop}$
outputs a score in [0,1], indicating the probability that a batch
has the property.  The adversary can use the average score across all
iterations to decide whether the target's entire dataset has the property
in question.

\subsection{Active property inference} 
\label{subsec:active-att}

An active adversary can perform a more powerful attack by using
\emph{multi-task learning}.  The adversary extends his local copy of
the collaboratively trained model with an augmented property classifier
connected to the last layer.  He trains this model to simultaneously
perform well on the main task and recognize batch properties.  On the
training data where each record has a main label $y$ and a property
label $p$, the model's joint loss is calculated as:
\begin{align*}
L_\text{mt} = \alpha\cdot L(x, y;\theta) + (1 - \alpha)\cdot L(x, p;
\theta)
\end{align*}
During collaborative training, the adversary uploads the updates
$\nabla_\theta L_\text{mt}$ based on this joint loss, causing the joint
model to learn separable representations for the data with and without
the property.  As a result, the gradients will be separable too (e.g.,
see~\Cref{fig:fb_active} in~\Cref{sec:active}), enabling the adversary
to tell if the training data has the property.

This adversary is still ``honest-but-curious'' in the cryptographic
parlance.  He faithfully follows the collaborative learning protocol and
does not submit any malformed messages.  The only difference with the
passive attack is that this adversary performs additional \emph{local}
computations and submits the resulting values into the collaborative
learning protocol.  Note that the ``honest-but-curious'' model does not
constrain the parties' input values, only their messages.

\section{Datasets and model architectures}

The datasets, collaborative learning tasks, and adversarial inference
tasks used in our experiments are reported in \Cref{tab:dataset_tasks}.
Our choices of hyperparameters are based on the standard models from
the ML literature.

\paragraphbe{Labeled Faces In the Wild (LFW).}  
LFW~\cite{huang2007labeled} contains 13,233 62x47 RGB face images for
5,749 individuals with labels such as gender, race, age, hair color,
and eyewear.

\paragraphbe{FaceScrub.} 
FaceScrub~\cite{ng2014data} contains 76,541 50x50 RGB images for 530
individuals with the gender label: 52.5\% are labeled as male, the rest
as female.  For our experiments, we selected a subset of 100 individuals
with the most images, for a total of 18,809 images.

\smallskip\noindent 
On both LFW and FaceScrub, the collaborative models are convolutional
neural networks (CNN) with three spatial convolution layers with 32,
64, and 128 filters, kernel size set to $(3,3)$, and max pooling layers
with pooling size set to 2, followed by two fully connected layers of
size 256 and 2.  We use rectified linear unit (ReLU) as the activation
function for all layers.  Batch size is 32 (except in the experiments
where we vary it), SGD learning rate is 0.01.

\paragraphbe{People in Photo Album (PIPA).} 
PIPA~\cite{zhang2015beyond} contains over 60,000 photos of 2,000
individuals collected from public Flickr photo albums.  Each image
includes one or more people and is labeled with the number of people and
their gender, age, and race.  For our experiments, we selected a subset
of 18,000 images with three or fewer people and scaled the raw images
to 128x128.

The collaborative model for PIPA is a VGG-style~\cite{simonyan2014very}
10-layer CNN with two convolution blocks consisting of one convolutional
layer and max pooling, followed by three convolution blocks consisting
of two convolutional layers and max pooling, followed by two fully
connected layers.  Batch size is 32, SGD learning rate is 0.01.

\paragraphbe{Yelp-health.} 
We extracted health care-related reviews from the Yelp
dataset\footnote{\url{https://www.yelp.com/dataset}} of 5 million reviews
of businesses tagged with numeric ratings (1-5) and attributes such
as business type and location.  Our subset contains 17,938 reviews
for 10 types of medical specialists.

\begin{table}[t]
\centering
\small
\setlength{\tabcolsep}{0.4em} 
\resizebox{\columnwidth}{!}{
\begin{tabular}{@{ }l|@{}r@{ }|l|l@{}}
\hline
{\bf Dataset} & {\bf ~\#Records} & {\bf Main tasks} & {\bf Inference tasks}\\
\hline\hline
{\bf LFW} & 13.2k & Gender/Smile/Age & Race/Eyewear\\
& & Eyewear/Race/Hair & \\ \hline
{\bf FaceScrub} & 18.8k & Gender & Identity \\ \hline
{\bf PIPA} & 18.0k & Age & Gender \\ \hline
{\bf CSI} & 1.4k  & Sentiment & Membership,\\ 
& & & Region/Gender/Veracity \\ \hline
{\bf FourSquare} & 15.5k & Gender & Membership\\ \hline
{\bf Yelp-health} & 17.9k & Review score & Membership,\\ 
& & & Doctor specialty \\ \hline
{\bf Yelp-author} & 16.2K  & Review score & Author\\
\hline 
\end{tabular}
}
\caption{Datasets and tasks used in our experiments.} %
\label{tab:dataset_tasks}
\vspace*{-0.2cm}
\end{table}

\paragraphbe{Yelp-author.} 
We also extracted a Yelp subset with the reviews of the top 10 most
prolific reviewers, 16,207 in total.

On both Yelp datasets, the model is a recurrent neural network with a
word-embedding layer of dimension 100.  Words in a review are mapped to
a sequence of word-embedding vectors, which is fed to a gated recurrent
unit (GRU~\cite{cho2014learning}) layer that maps it to a sequence of
hidden vectors.  We add a fully connected classification layer to the
last hidden vector of the sequence.  SGD learning rate is 0.05.

\paragraphbe{FourSquare.} 
In~\cite{yang2015nationtelescope,yang2015participatory}, Yang et
al.\ collected a global dataset of FourSquare location ``check-ins''
(userID, time, location, activity) from April 2012 to September 2013.
For our experiments, we selected a subset of 15,548 users who checked in
at least 10 different locations in New York City and for whom we know
their gender~\cite{yang2016privcheck}.  This yields 528,878 check-ins.
The model is a gender classifier, a task previously studied by Pang et
al.~\cite{pang2016deepcity} on similar datasets.

\paragraphbe{CLiPS Stylometry Investigation (CSI) Corpus.} 
This annually expanded dataset~\cite{verhoeven2014clips} contains
student-written essays and reviews.  We obtained 1,412 reviews, equally
split between Truthful/Deceptive or Positive/Negative and labeled with
attributes of the author (gender, age, sexual orientation, region of
origin, personality profile) and the document (timestamp, genre, topic,
veracity, sentiment).  80\% of the reviews are written by females, 66\%
by authors from Antwerpen, the rest from other parts of Belgium and
the Netherlands.

\smallskip\noindent 
On both the FourSquare and CSI datasets, the model, which is based
on~\cite{kim2014convolutional}, first uses an embedding layer to turn
non-negative integers (locations indices and word tokens) into dense
vectors of dimension 320, then applies three spatial convolutional layers
with 100 filters and variable kernel windows of size $(3,320)$, $(4,320)$
and $(5,320)$ and max pooling layers with pooling size set to $(l-3,
1)$, $(l-4,1)$, and $(l-5,1)$ where $l$ is the fixed length to which
input sequences are padded.  The hyperparameter $l$ is 300 on CSI and
100 on FourSquare.  After this, the model has two fully connected layers
of size 128 and 2 for FourSquare and one fully connected layer of size
2 for CSI.  We use RELU as the activation function.  Batch size is 100
for FourSquare, 12 for CSI.  SGD learning rate is 0.01.

\section{Two-Party Experiments}
\label{sec:experiments2}

All experiments were performed on a workstation running Ubuntu
Server 16.04 LTS equipped with a 3.4GHz CPU i7-6800K, 32GB RAM,
and an NVIDIA TitanX GPU card.  We use MxNet~\cite{chen2015mxnet}
and Lasagne~\cite{lasagne} to implement deep neural networks and
Scikit-learn~\cite{scikit-learn} for conventional ML models. The source
code is available upon request.  Training our inference models takes
less than 60 seconds on average and does not require a GPU.

We use AUC scores to evaluate the performance of both the
collaborative model and our property inference attacks.  For membership
inference, we report only precision because our decision rule from
Section~\ref{subsec:membership-att} is binary and does not produce a
probability score.

\begin{table}[t]
\centering
\resizebox{0.7\columnwidth}{!}{
\begin{tabular}{rr|rr}
\hline
\multicolumn{2}{c|}{\textbf{Yelp-health}} & \multicolumn{2}{c}{\textbf{FourSquare}} \\
\hline
{\bf Batch Size} & {\bf Precision} & {\bf Batch Size} & {\bf Precision}
\\ \hline\hline
32 & 0.92 & 100 & 0.99 \\ 
64 & 0.84 & 200 & 0.98  \\ 
128 & 0.75 & 500 & 0.91  \\ 
256 & 0.66 & 1,000 & 0.76 \\ 
512 & 0.62 & 2,000 & 0.62  \\
\hline 
\end{tabular}}
\caption{Precision of membership inference (recall is 1).}
\label{tab:location_mia}
\vspace*{-0cm}
\end{table}

\subsection{Membership inference}
\label{sec:two-member}

The adversary first builds a Bag of Words (BoW) representation for the
input whose membership in the target's training data he aims to infer.
We denote this as the test BoW.  During training, as explained in
Section~\ref{subsec:membership-att}, the non-zero gradients of the
embedding layer reveal which ``words'' are present in each batch of the
target's data, enabling the adversary to build a batch BoW.  If the test
BoW is a subset of the batch BoW, the adversary infers that the input
of interest occurs in the batch.

We evaluate membership inference on the Yelp-health and FourSquare
datasets with the vocabulary of 5,000 most frequent words and 30,000
most popular locations, respectively.  We split the data evenly between
the target and the adversary and train a collaborative model for 3,000
iterations.

Table~\ref{tab:location_mia} shows the precision of membership inference
for different batch sizes.  As batch size increases, the adversary
observes more words in each batch BoW and the attack produces more
false positives.  Recall is always perfect (i.e., no false negatives)
because any true test BoW must be contained in at least one of the batch
BoWs observed by the adversary.

\subsection{Single-batch property inference}
\label{ssec:2p_batch}

We call a training batch $\bnonprop$ if none of the inputs in it have
the property, $\bprop$ otherwise.  The adversary aims to identify which
of the batches are $\bprop$.  We split the training data evenly between
the target and the adversary and assume that the same fraction of inputs
in both subsets have the property.  During training, $\frac{1}{m}$
of the target's batches include only inputs with the property ($m=2$
in the following).

\begin{table}[t]
\small
\centering
\setlength{\tabcolsep}{0.4em} 
\resizebox{\columnwidth}{!}{
\begin{tabular}{l|l|r|r||l|l|r|r}
\hline
{\bf Main T.} & {\bf Infer T.} & {\bf Corr.} & {\bf AUC} & {\bf Main T.} & {\bf Infer T.} & {\bf Corr.} & {\bf AUC} \\ \hline\hline
Gender & Black & -0.005 & 1.0 & Gender & Sunglasses & -0.025 & 1.0 \\
Gender & Asian & -0.018 & 0.93 & Gender & Eyeglasses & 0.157 & 0.94 \\ \hline 
Smile & Black & 0.062 & 1.0 & Smile & Sunglasses & -0.016 & 1.0 \\
Smile & Asian & 0.047 & 0.93 & Smile & Eyeglasses & -0.083 & 0.97 \\ \hline 
Age & Black & -0.084 & 1.0 & Race & Sunglasses & 0.026 & 1.0 \\
Age & Asian & -0.078 & 0.97 & Race & Eyeglasses & -0.116 & 0.96 \\ \hline 
Eyewear & Black & 0.034 & 1.0 & Hair & Sunglasses & -0.013 & 1.0 \\
Eyewear & Asian & -0.119 & 0.91 & Hair & Eyeglasses & 0.139 & 0.96 \\ \hline 
\end{tabular}
}
\caption{AUC score of single-batch property inference on LFW.  We also
report the Pearson correlation between the main task label and the
property label.}
\label{tbl:lfw_prop} 
\vspace*{-0.0cm}
\end{table}

\begin{figure*}[t]
\centering
\begin{subfigure}[b]{0.255\textwidth}
\includegraphics[width=1\textwidth]{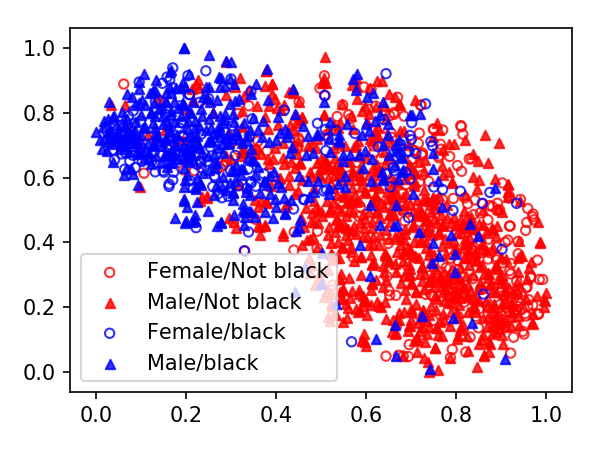}
\caption{\rm pool1}
\end{subfigure}
\hspace*{-0.35cm}
\begin{subfigure}[b]{0.255\textwidth}
\includegraphics[width=1\textwidth]{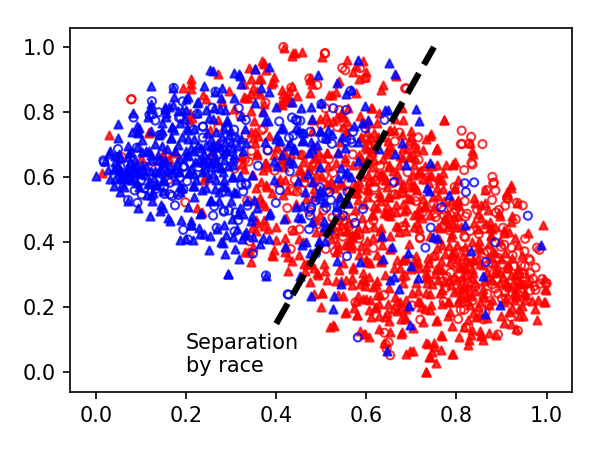}
\caption{\rm pool2}
\end{subfigure}
\hspace*{-0.35cm}
\begin{subfigure}[b]{0.255\textwidth}
\includegraphics[width=1\textwidth]{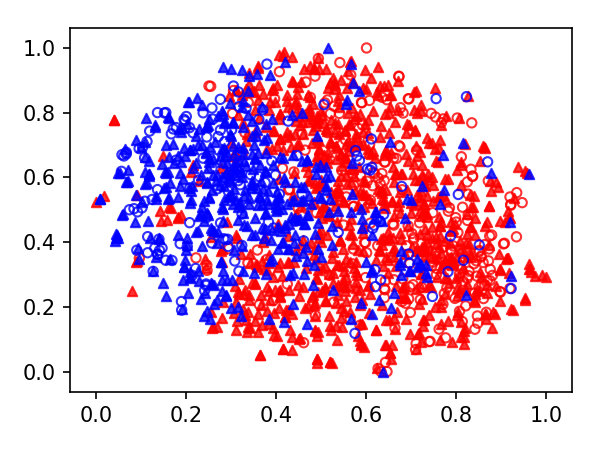}
\caption{\rm pool3}
\end{subfigure}
\hspace*{-0.35cm}
\begin{subfigure}[b]{0.255\textwidth}
\includegraphics[width=1\textwidth]{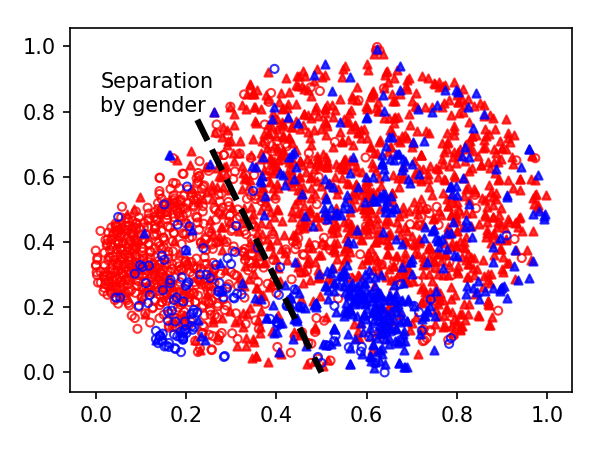}
\caption{\rm fc}
\end{subfigure}
\caption{t-SNE projection of the features from different layers of
the joint model on LFW gender classification; hollow circle point is female, solid triangle point is male, blue point is the property ``race: black'' and red point is data without the property. }
\label{fig:lfw_tsne}
\vspace*{-0.15cm}
\end{figure*}

\begin{figure*}[t]
\centering
\begin{subfigure}[b]{0.36\textwidth}
\includegraphics[width=1\textwidth]{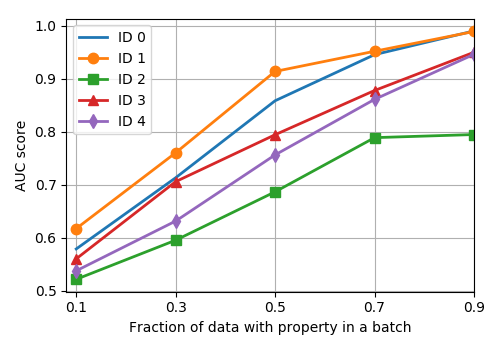} 
\caption{\rm FaceScrub}
\end{subfigure}
\begin{subfigure}[b]{0.36\textwidth}
\includegraphics[width=1\textwidth]{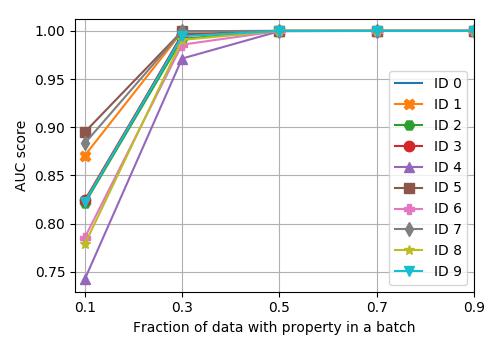} 
\caption{\rm Yelp-author}
\end{subfigure}
\caption{AUC vs.\ the fraction of the batch that has the property on
FaceScrub and Yelp-author.}
\label{fig:fb_fraction}
\vspace*{-0.15cm}
\end{figure*}

\paragraphbe{LFW.} 
Table~\ref{tbl:lfw_prop} reports the results of single-batch
property inference on the LFW dataset.  We chose properties that
are \emph{uncorrelated} with the main classification label that the
collaborative model is trying to learn.  The attack has perfect AUC
when the main task is gender classification and the inference task is
``race:black'' (the Pearson correlation between these labels is -0.005).
The attack also achieves almost perfect AUC when the main task is
``race: black'' and the inference task is ``eyewear: sunglasses.''
It also performs well on several other properties, including ``eyewear:
glasses'' when the main task is ``race: Asian.''

These results demonstrate that gradients observed during training
leak more than the characteristic features of each class.  In fact,
\textbf{collaborative learning leaks properties of the training data
that are uncorrelated with class membership}.  To understand why, we
plot the t-SNE projection~\cite{maaten2008visualizing} of the features
from different layers of the joint model in~\Cref{fig:lfw_tsne}.
Observe that the feature vectors are grouped by property in the lower
layers pool1, pool2 and pool3, and by class label in the higher layer.
Intuitively, the model did not just learn to separate inputs by class.
The lower layers of the model also learned to separate inputs by various
properties that are uncorrelated with the model's designated task.
Our inference attack exploits this unintended extra functionality.

\paragraphbe{Yelp-health.}  
\label{yelpattack}
On this dataset, we use review-score classification as the main task and
the specialty of the doctor being reviewed as the property inference task.
Obviously, the latter is more sensitive from the privacy perspective.

We use 3,000 most frequent words in the corpus as the vocabulary and train
for 3,000 iterations.  Using BoWs from the embedding-layer gradients, the
attack achieves almost perfect AUC.  Table~\ref{tbl:yelp_words} shows the
words that have the highest predictive power in our logistic regression.

\begin{table}[t]
\small
\centering
\setlength{\tabcolsep}{0.4em} 
\resizebox{\columnwidth}{!}{
\begin{tabular}{l|l}
\hline
{\bf Health Service} & {\bf Top Words in Positive Class}  \\ \hline \hline
Obstetricians & pregnancy, delivery, women, birth, ultrasound \\
Pediatricians & pediatrics, sick, parents, kid, newborn \\
Cosmetic Surgeons & augmentation, plastic, breast, facial, implants\\
Cardiologists & cardiologist, monitor, bed, heart, ER \\
Dermatologists & acne, dermatologists, mole, cancer, spots \\
Ophthalmologists & vision, LASIK, contacts, lenses, frames\\
Orthopedists & knee, orthopedic, shoulder, injury, therapy\\
Radiologists & imaging, SimonMed, mammogram, CT, MRI\\
Psychiatrists & psychiatrist, mental, Zedek, depression, sessions\\
Urologists & Edgepark, pump, supplies, urologist, kidney \\
\hline
\end{tabular}}
\caption{Words with the largest positive coefficients in 
the property classifier for Yelp-health.}
\label{tbl:yelp_words}
\end{table}

\paragraphbe{Fractional properties.} 
We now attempt to infer that \emph{some} of the inputs in a batch have
the property.  For these experiments, we use FaceScrub's top 5 face
IDs and Yelp-author (the latter with the 3,000 most frequent words as
the vocabulary).  The model is trained for 3,000 iterations.  As before,
$1/2$ of the target's batches include inputs with the property, but here
we vary the fraction of inputs with the property within each such batch
among 0.1, 0.3, 0.5, 0.7, and 0.9.

\Cref{fig:fb_fraction} reports the results.  On FaceScrub for IDs 0, 1,
and 3, AUC scores are above 0.8 even if only 50\% of the batch contain
that face, i.e., the adversary can successfully infer that photos of a
particular person appear in a batch even though (a) the model is trained
for generic gender classification, and (b) half of the photos in the
batch are of other people.  If the fraction is higher, AUC approaches 1.

On Yelp-author, AUC scores are above 0.95 for all identities even when
the fraction is 0.3, i.e., the attack successfully infers the authors
of reviews even though (a) the model is trained for generic sentiment
analysis, and (b) more than two thirds of the reviews in the batch are
from other authors.

\begin{figure*}[t]
\centering
\begin{subfigure}[b]{0.43\textwidth}
\includegraphics[width=1\textwidth]{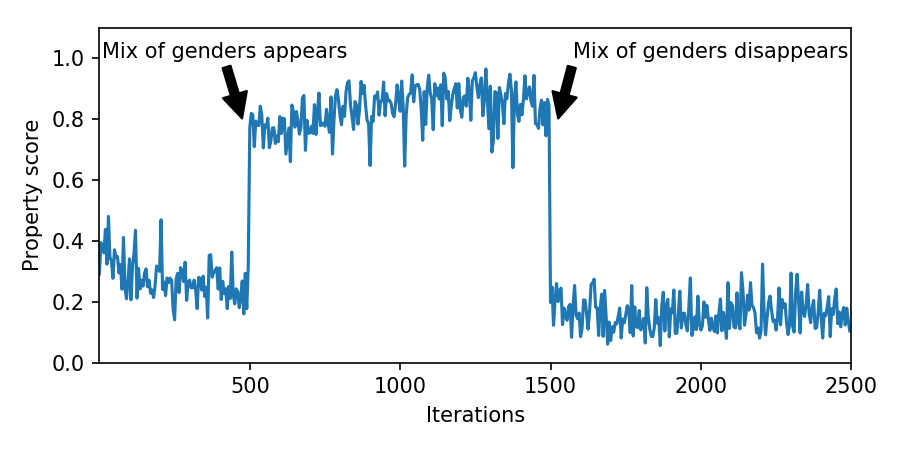}
\caption{\rm PIPA}
\label{fig:fb_it-a}
\end{subfigure}
\begin{subfigure}[b]{0.43\textwidth}
\includegraphics[width=1\textwidth]{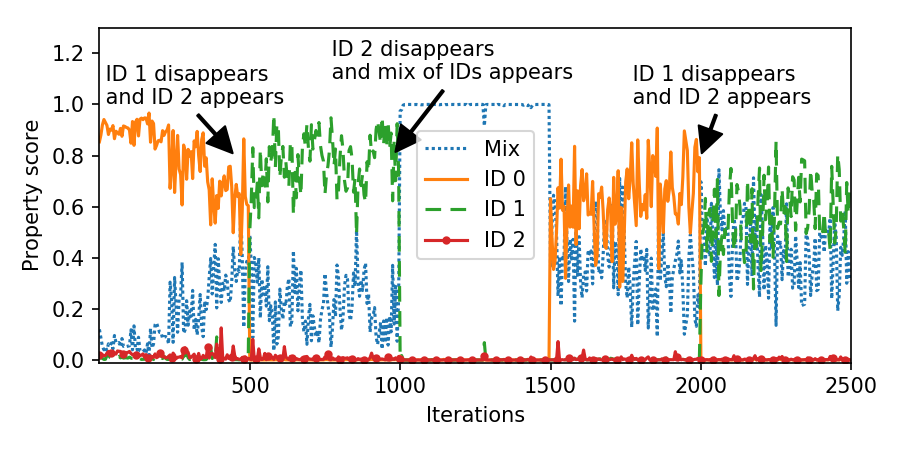}
\caption{\rm FaceScrub}
\label{fig:fb_it-b}
\end{subfigure}
\caption{Inferring occurrence of a single-batch property.}
\label{fig:fb_it}
\vspace*{-0.15cm}
\end{figure*}

\subsection{Inferring when a property occurs} 
\label{sec:emerge}

Continuous training, when new training data is added to the process as
it becomes available, presents interesting opportunities for inference
attacks.  If the occurrences of a property in the training data can
be linked to events outside the training process, privacy violation
is exacerbated.  For example, suppose a model leaks that a certain
third person started appearing in another participant's training data
immediately after that participant uploaded his photos from a trip.

\paragraphbe{PIPA.} 
Images in the PIPA dataset have between 1 to 3 faces.  We train the
collaborative model to detect if there is a young adult in the image;
the adversary's inference task is to determine if people in the image are
of the same gender.  The latter property is a stepping stone to inferring
social relationships and thus sensitive.  We train the model for 2,500
iterations and let the batches with the ``same gender'' property appear
in iterations 500 to 1500.

\Cref{fig:fb_it-a} shows, for each iteration, the probability output
by the adversary's classifier that the batch in that iteration has
the property.  The appearance and disappearance of the property in the
training data are clearly visible in the plot.

\paragraphbe{FaceScrub.} 
For the gender classification model on FaceScrub, the adversary's
objective is to infer whether and when a certain person appears in
the other participant's photos.  The joint model is trained for 2,500
iterations.  We arrange the target's training data so that two specific
identities appear during certain iterations: ID 0 in iterations 0 to
500 and 1500 to 2000, ID 1 in iterations 500 to 1000 and 2000 to 2500.
The rest of the batches are mixtures of other identities.  The adversary
trains three property classifiers, for ID 0, ID 1, and also for ID 2
which does not appear in the target's dataset.

\Cref{fig:fb_it-b} reports the scores of all three classifiers.  ID 0
and 1 receive the highest scores in the iterations where they appear,
whereas ID 2, which never appears in the training data, receives very
low scores in all iterations.

These experiments show that our attacks can successfully infer
\emph{dynamic properties} of the training dataset as collaborative
learning progresses.

\subsection{Inference against well-generalized models}

\begin{figure}[t]
\centering
\includegraphics[width=0.36 \textwidth]{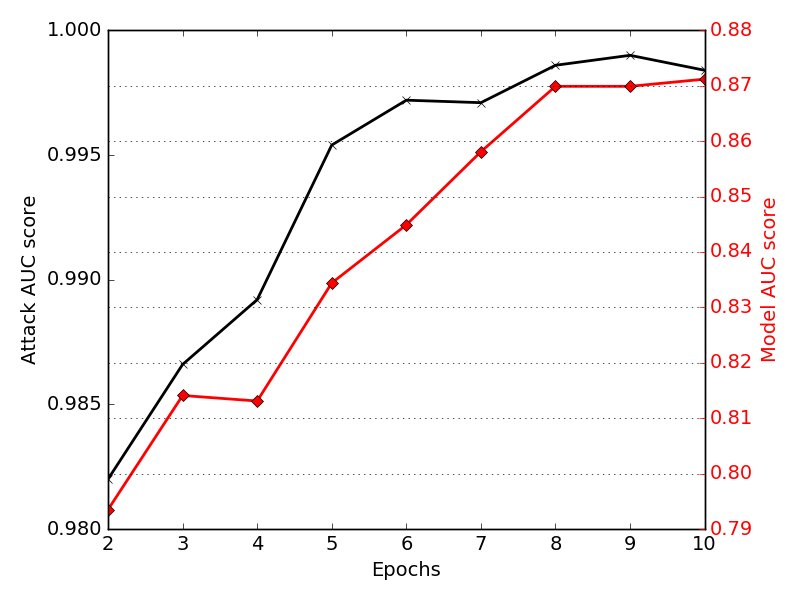}
\caption{Attack performance with respect to the number of collaborative
learning epochs. }
\label{fig:csi_epochs}
\vspace*{-0.2cm}
\end{figure}

To show that our attacks work with (1) relatively few observed model
updates and (2) against well-generalized models, we experiment with the
CSI corpus.  \Cref{fig:csi_epochs} reports the accuracy of inferring
the author's gender.  The attack reaches $0.98$ AUC after only 2 epochs
and improves as the training progresses and the adversary collects more
updates.

\Cref{fig:csi_epochs} also shows that the model is not overfitted.
Its test accuracy on the main sentiment-analysis task is high and improves
with the number of the epochs.

\begin{figure*}[t]
\centering
\begin{subfigure}[b]{0.32\textwidth}
\includegraphics[width=0.95\textwidth]{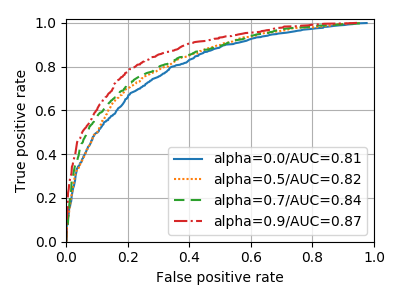}
\caption{\rm ROC for different $\alpha$}
\label{fig:fb_active-a}
\end{subfigure}
\begin{subfigure}[b]{0.32\textwidth}
\includegraphics[width=0.95\textwidth]{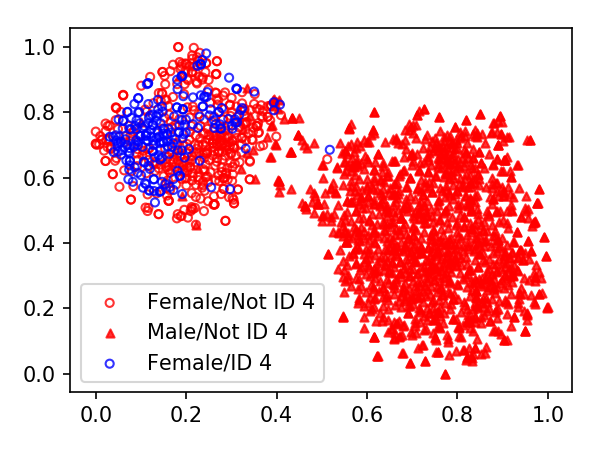}
\caption{\rm t-SNE of the final layer for $\alpha=0$}
\label{fig:fb_active-b}
\end{subfigure}
\begin{subfigure}[b]{0.32\textwidth}
\includegraphics[width=0.95\textwidth]{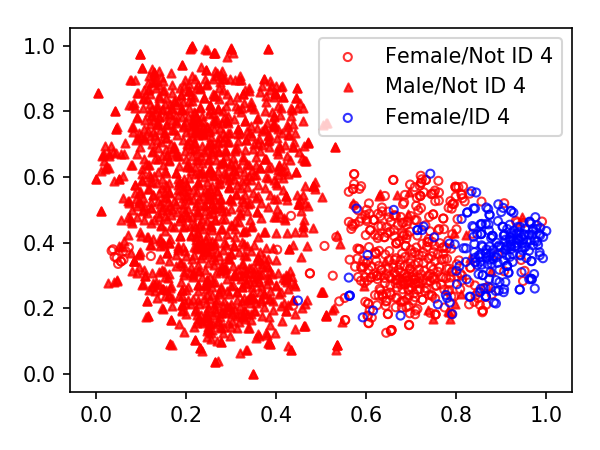}
\caption{\rm  t-SNE of the final layer for $\alpha=0.7$. }
\label{fig:fb_active-c}
\end{subfigure}
\caption{Two-party active property inference attack on FaceScrub. For (b) and (c),  hollow circle point is female, solid triangle point is male, blue point is the property ``ID 4'' and red point is data without the property.}
\label{fig:fb_active}
\vspace*{-0.3cm}
\end{figure*}

\subsection{Active property inference}
\label{sec:active}

To show the additional power of the active attack
from~\Cref{subsec:active-att}, we use FaceScrub.  The main task is gender
classification, the adversary's task is to infer the presence of ID 4
in the training data.  We assume that this ID occurs in a single batch,
where it constitutes 50\% of the photos.  We evaluate the attack with
different choices of $\alpha$, which controls the balance between the
main-task loss and the property-classification loss in the adversary's
objective function.

\Cref{fig:fb_active-a} shows that AUC increases as we increase $\alpha$.
\Cref{fig:fb_active-b} and \Cref{fig:fb_active-c} show the t-SNE
projection of the final fully connected layer, with $\alpha=0$ and
$\alpha=0.7$, respectively.  Observe that the data with the property
(blue points) is grouped tighter when $\alpha=0.7$ than in the model
trained under a passive attack ($\alpha=0$).  This illustrates that
\emph{as a result of the active attack, the joint model learns a better
separation for data with and without the property}.

\section{Multi-Party Experiments}
\label{sec:experimentsM}

In the multi-party setting, we only consider passive property inference
attacks.  We vary the number of participants between 4 and 30 to match
the deployment scenarios and applications proposed for collaborative
learning, e.g., hospitals or biomedical research institutions training
on private medical data~\cite{jochems1,jochems2}.  This is similar
to prior work~\cite{hitaj2017deep}, which was evaluated on MNIST
with 2 participants and face recognition on the AT\&T dataset with
41 participants.

\begin{figure*}
\centering
\begin{subfigure}[b]{0.36\textwidth}
\includegraphics[width=1\textwidth]{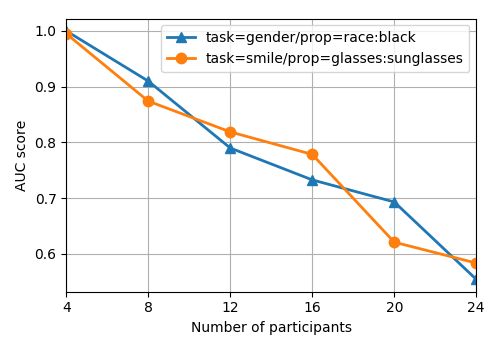}
\caption{\rm LFW}
\label{fig:multi_sgd-a}
\end{subfigure}
~
\begin{subfigure}[b]{0.36\textwidth}
\includegraphics[width=1\textwidth]{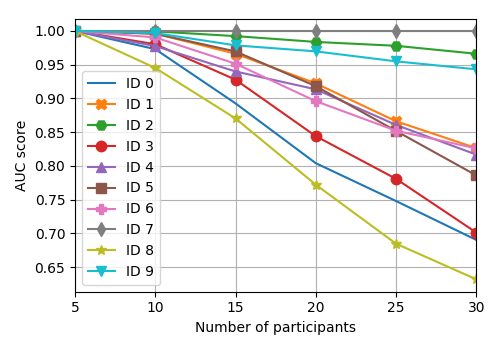}
\caption{\rm Yelp-author}
\label{fig:multi_sgd-b}
\end{subfigure}
\caption{Multi-party learning with synchronized SGD: attack AUC score vs.\
the number of participants.}
\label{fig:multi_sgd}
\vspace*{-0.15cm}
\end{figure*}

\subsection{Synchronized SGD}
\label{sec:multi-sgd}
\label{aggregate-attacks}

As the number of honest participants in collaborative learning increases,
the adversary's task becomes harder because the observed gradient updates
are aggregated across multiple participants.  Furthermore, the inferred
information may not directly reveal the identity of the participant to
whom the data belongs (see Section~\ref{sec:attribution}).

In the following experiments, we split the training data evenly
across all participants, but so that only the target and the
adversary have the data with the property.  The joint model is
trained with the same hyperparameters as in the two-party case.
Similar to~\Cref{ssec:2p_batch}, the adversary's goal is to identify
which aggregated gradient updates are based on batches $\bprop$ with
the property.

\paragraphbe{LFW.} 
We experiment with (1) gender classification as the main task and
``race: black'' as the inference task, and (2) smile classification
as the main task and ``eyewear: sunglasses'' as the inference task.
\Cref{fig:multi_sgd-a} shows that the attack still achieves reasonably
high performance, with AUC score around 0.8, when the number of
participants is 12.  Performance then degrades for both tasks.

\paragraphbe{Yelp-author.} 
The inference task is again author identification.  In the multi-party
case, the gradients of the embedding layer leak the batch BoWs of
all honest participants, not just the target.  \Cref{fig:multi_sgd-b}
reports the results.  For some authors, AUC scores do not degrade
significantly even with many participants.  This is likely due to some
unique combinations of words used by these authors, which identify them
even in multi-party settings.

\begin{figure*}[ht]
\centering
\begin{subfigure}[b]{0.2655\textwidth}
\includegraphics[width=0.95\textwidth]{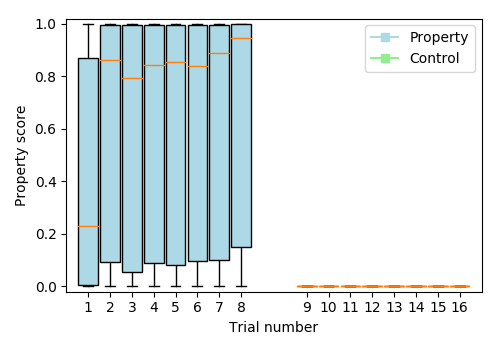}
\caption{\rm Face ID 1, $K=3$}
\end{subfigure}
\hspace*{-0.6cm}
\begin{subfigure}[b]{0.2655\textwidth}
\includegraphics[width=0.95\textwidth]{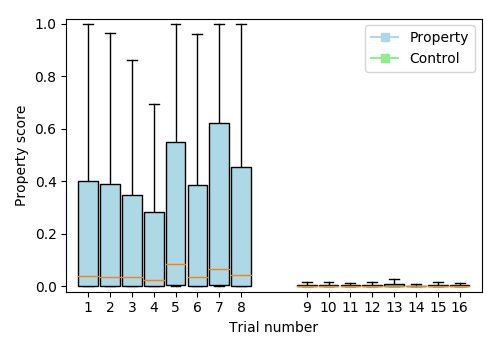}
\caption{\rm Face ID 1, $K=5$}
\end{subfigure}
\hspace*{-0.6cm}
\begin{subfigure}[b]{0.2655\textwidth}
\includegraphics[width=0.95\textwidth]{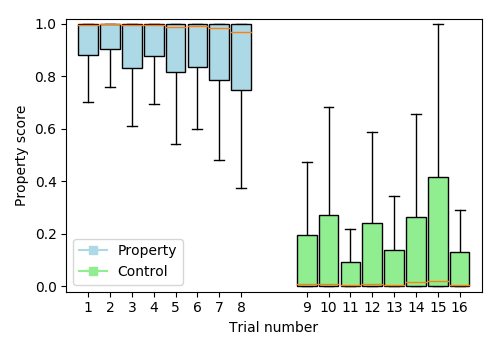}
\caption{\rm  Face ID 3, $K=3$}
\end{subfigure}
\hspace*{-0.6cm}
\begin{subfigure}[b]{0.2655\textwidth}
\includegraphics[width=0.95\textwidth]{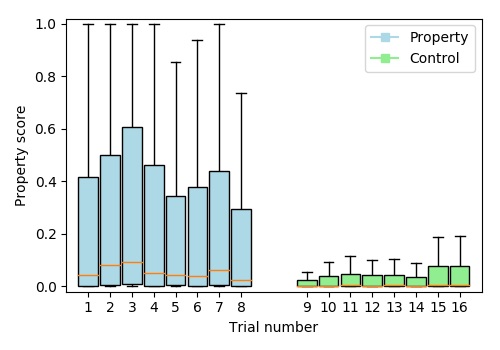}
\caption{\rm  Face ID 3, $K=5$}
\end{subfigure}
\caption{Multi-party learning with model averaging.  Box plots show the
distribution of the adversary's scores in each trial: in the 8 trials on
the left, one participant's data has the property; in the 8 trials on the
right, none of the honest participants have the data with the property.}
\label{fig:multi_ma}
\vspace*{-0.3cm}
\end{figure*}

\subsection{Model averaging}
\label{sec:multi-avg}

In every round $t$ of federated learning with model averaging (see
Algorithm~\ref{alg:fl}), the adversary observes $\theta_t-\theta_{t-1} =
\sum_{k}\frac{n^k}{n}\theta_t^k - \sum_{k}\frac{n^k}{n}\theta_{t-1}^k =
\sum_{k}\frac{n^k}{n} (\theta_t^k - \theta_{t-1}^k)$, where $\theta_t^k
- \theta_{t-1}^k$ are the aggregated gradients computed on the $k$-th
participant's local dataset.

In our experiments, we split the training data evenly among honest
participants but ensure that in the target participant's subset,
$\hat{p}$\% of the inputs have the property, while none of the other
honest participants' data have it.  During each epoch of local training,
every honest participant splits his local training data into 10 batches
and performs one round of training.

We assume that the adversary has the same number of inputs with the
property as the target.  As before, when the adversary trains his binary
classifier, he needs to locally ``emulate'' the collaborative training
process, i.e., sample data from his local dataset, compute aggregated
updates, and learn to distinguish between the aggregates based on the
data without the property and aggregates where one of the underlying
updates was based on the data with the property.

We perform 8 trials where a subset of the training data has the property
and 8 control trials where there are no training inputs with the property.

\paragraphbe{Inferring presence of a face.} 
We use FaceScrub and select two face IDs (1 and 3) whose presence we
want to infer.  In the ``property'' case, $\hat{p}=80\%$, i.e., 80\%
of one honest participant's training data consist of the photos that
depict the person in question.  In the control case, $\hat{p}=0\%$,
i.e., the photos of this person do not occur in the training data.
\Cref{fig:multi_ma} shows the scores assigned by the adversary's
classifier to the aggregated updates with 3 and 5 total participants.
When the face in question is present in the training dataset, the scores
are much higher than when it is absent.

Success of the attack depends on the property being inferred, distribution
of the data across participants, and other factors.  For example, the
classifiers for Face IDs 2 and 4, which were trained in the same fashion
as the classifiers for Face IDs 1 and 3, failed to infer the presence
of the corresponding faces in the training data.

\begin{figure}
\centering
\includegraphics[width=0.4\textwidth]{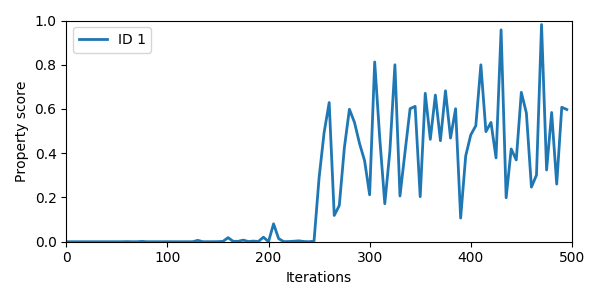}
\vspace*{-0.15cm}
\caption{Inferring that a participant whose local data has the property
of interest has joined the training.  $K=2$ for rounds 0 to 250, $K=3$
for rounds 250 to 500.}
\label{fig:ma_it}
\end{figure}

\paragraphbe{Inferring when a face occurs.} 
In this experiment, we aim to infer when a participant whose local data
has a certain property joined collaborative training.  We first let
the adversary and the rest of the honest participants train the joint
model for 250 rounds.  The target participant then joins the training at
round $t=250$ with the local data that consists of photos depicting ID 1.
\Cref{fig:ma_it} reports the results of the experiment: the adversary's
AUC scores are around 0 when face ID 1 is not present and then increase
almost to 1.0 right after the target participant joins the training.

\section{Defenses}
\label{sec:defenses}

\subsection{Sharing fewer gradients} 
\label{sec:grad_frac}

As suggested in~\cite{shokri2015privacy}, participants in collaborative
learning could share only a fraction of their gradients during each
update.  This reduces communication overhead and, potentially, leakage,
since the adversary observes fewer gradients.

To evaluate this defense, we measure the performance of single-batch
inference against a sentiment classifier collaboratively trained on
the CSI Corpus by two parties who exchange only a fraction of their
gradients.  \Cref{tab:grad_frac_defense} shows the resulting AUC scores:
when inferring the region of the texts' authors, our attack still achieves
0.84 AUC when only 10\% of the updates are shared during each iteration,
compared to 0.93 AUC when all updates are shared.

\subsection{Dimensionality reduction}

\begin{table}[t]
\centering
\small
\resizebox{0.85\columnwidth}{!}{
\begin{tabular}{l|r|r|r}
\hline
{\bf Property / \% parameters update} & {\bf 10\%} & {\bf 50\%} & {\bf 100\%}
\\ \hline\hline
Top region (Antwerpen) & 0.84 & 0.86 & 0.93 
\\ 
Gender & 0.90 & 0.91 & 0.93
\\ 
Veracity & 0.94 & 0.99  & 0.99 
\\ \hline
\end{tabular}
}
\caption{Inference attacks against the CSI Corpus for different
fractions of gradients shared during training.}
\label{tab:grad_frac_defense}
\vspace*{-0.3cm}
\end{table}

\begin{figure}[t]
\centering
\includegraphics[width=0.35\textwidth]{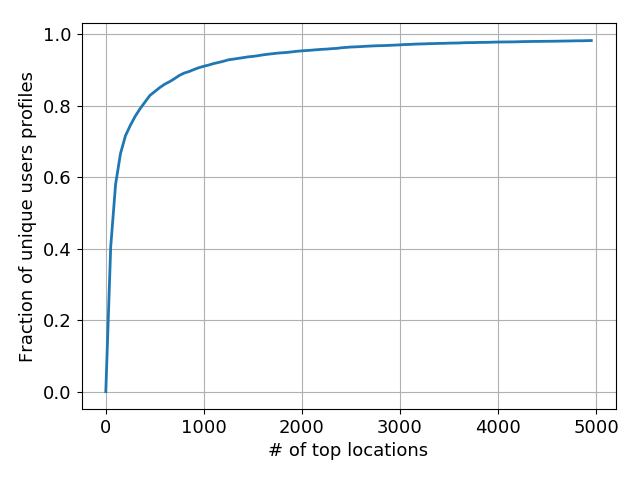}
\vspace*{-0.2cm}
\caption{Uniqueness of user profiles with respect to the number of
top locations.}
\label{fig:uniqueness}
\vspace*{-0.1cm}
\end{figure}

As discussed in Section~\ref{sec:embedleak}, if the input space of the
model is sparse and inputs must be embedded into a lower-dimensional
space, non-zero gradient updates in the embedding layer reveal which
inputs are present in the training batch.

One plausible defense is to only use inputs that occur many times in the
training data.  This does not work in general, e.g., \Cref{fig:uniqueness}
shows that restricting inputs to the top locations in the FourSquare
dataset eliminates most of the training data.

A smarter defense is to restrict the model so that it only uses ``words''
from a pre-defined vocabulary of common words.  For example, Google's
federated learning for predictive keyboards uses a fixed vocabulary of
5,000 words~\cite{mcmahan2016communication}.  In \Cref{tab:nlp_defense},
we report the accuracy of our membership inference attack and the accuracy
of the joint model on its main task\textemdash gender classification for
the FourSquare dataset, sentiment analysis for the CSI Corpus\textemdash
for different sizes of the common vocabulary (locations and words,
respectively).  This approach partially mitigates our attacks but also
has a significant negative impact on the quality of the collaboratively
trained models.

\begin{table}[t]
\centering
\small
\resizebox{0.95\columnwidth}{!}{
\begin{tabular}{rrr|rrr}
\multicolumn{3}{c|}{\textbf{CSI}}  & \multicolumn{3}{c}{\textbf{FourSquare}} \\
\hline
{\bf Top N} & {\bf Attack} & {\bf Model} & {\bf Top N} & {\bf Attack} & {\bf Model} \\
{\bf words} &  {\bf Precision} & {\bf AUC} & {\bf locations} & {\bf Precision} & {\bf AUC}
\\ \hline\hline
4,000 & 0.94 & 0.91 & 30,000 & 0.91 & 0.64 \\ 
2,000 & 0.92 & 0.87 & 10,000 & 0.86 & 0.59 \\ 
1,000 & 0.92 & 0.85 & 3,000 & 0.65 & 0.51  \\ 
500 & 0.82 & 0.84 & 1,000 & 0.52 & 0.50\\ \hline
\end{tabular}}
\caption{Membership inference against the CSI Corpus and FourSquare for
different vocabulary sizes.}
\label{tab:nlp_defense}
\vspace*{-0.3cm}
\end{table}

\subsection{Dropout}

Another possible defense is to employ
\emph{dropout}~\cite{srivastava2014dropout}, a popular regularization
technique used to mitigate overfitting in neural networks.  Dropout
randomly deactivates activations between neurons, with probability
$p_{drop} \in [0,1]$.  Random deactivations may weaken our attacks
because the adversary observes fewer gradients corresponding to the
active neurons.

To evaluate this approach, we add dropout after the max pool layers in
the joint model.  \Cref{tab:dropout_defense} reports the accuracy of
inferring the region of the reviews in the CSI Corpus, for different
values of $p_{drop}$.  Increasing the randomness of dropout makes our
attacks \emph{stronger} while slightly decreasing the accuracy of
the joint model.  Dropout stochastically removes features at every
collaborative training step, thus yielding more \emph{informative}
features (similar to feature bagging~\cite{ho1995random,chang2017dropout})
and increasing variance between participants' updates.

\subsection{Participant-level differential privacy}
\label{ssec:dp_defense}

As discussed in Section~\ref{sec:collabml}, record-level
$\varepsilon$-differential privacy, by definition, bounds the success
of membership inference but does not prevent property inference.
Any application of differential privacy entails application-specific
tradeoffs between privacy of the training data and accuracy of the
resulting model.  The participants must also somehow choose the parameters
(e.g., $\varepsilon$) that control this tradeoff.

In theory, participant-level differential privacy bounds the success
of inference attacks described in this paper.  We implemented the
participant-level differentially private federated learning algorithm
by McMahan et al.~\cite{mcmahan2017learning} and attempted to train a 
gender classifier on LFW, but the model did not converge for any number 
of participants (we tried at most 30).
This is due to the magnitude of noise needed to achieve differential
privacy with the moments accountant bound~\cite{abadi2016deep},
which is inversely proportional to the number of users (the model
in~\cite{mcmahan2017learning} was trained on \emph{thousands} of users).
Another participant-level differential privacy mechanism, presented
in~\cite{geyer2017differentially}, also requires a very large number
of participants.  Moreover, these two mechanisms have been used,
respectively, for language modeling~\cite{mcmahan2017learning} and
handwritten digit recognition~\cite{geyer2017differentially}.  Adapting
them to the specific models and tasks considered in this paper may not
be straightforward.

Following~\cite{mcmahan2017learning, geyer2017differentially}, we believe
that participant-level differential privacy provide reasonable accuracy
only in settings involving at least thousands of participants.  We believe
that further work is needed to investigate whether participant-level
differential privacy can be adapted to prevent our inference attacks
\emph{and} obtain high-quality models in settings that do not involve
thousands of users.

\begin{table}[t]
\centering
\small
\resizebox{0.75\columnwidth}{!}{
\begin{tabular}{r|r|r}
\hline
{\bf Dropout Prob.} & {\bf Attack AUC} & {\bf Model AUC}
\\ \hline\hline
0.1 & 0.94 & 0.87 \\ 
0.3 & 0.97 & 0.87 \\
0.5 & 0.98 & 0.87 \\ 
0.7 & 0.99 & 0.86 \\
0.9 & 0.99 & 0.84 \\ \hline
\end{tabular}}
\caption{Inference of the top region (Antwerpen) on the CSI Corpus for
different values of dropout probability.}
\label{tab:dropout_defense}
\vspace*{-0.1cm}
\end{table}

\section{Limitations of the attacks}

\subsection{Auxiliary data}

Our property inference attacks assume that the adversary has auxiliary
training data correctly labeled with the property he wants to infer.
For generic properties, such data is easy to find.  For example,
the auxiliary data for inferring the number and genders of people can
be any large dataset of images with males and females, single and in
groups, where each image is labeled with the number of people in it
and their genders.  Similarly, the auxiliary data for inferring the
medical specialty of doctors can consist of any texts that include words
characteristic of different specialties (see Table~\ref{tbl:yelp_words}).

More targeted inference attacks require specialized auxiliary data that
may not be available to the adversary.  For example, to infer that photos
of a certain person occurs in another participant's dataset, the adversary
needs (possibly different) photos of that person to train on.  To infer
the authorship of training texts, the adversary needs a sufficiently
large sample of texts known to be written by a particular author.

\subsection{Number of participants} 

In our experiments, the number of participants in collaborative
training is relatively small (ranging from 2 to 30), while some
federated-learning applications involve thousands or millions
of users~\cite{mcmahan2016communication, mcmahan2017learning}.
As discussed in Section~\ref{sec:multi-sgd}, performance of our attacks
drops significantly as the number of participants increases.

\subsection{Undetectable properties} 

It may not be possible to infer some properties from model updates.  For
example, our attack did not detect the presence of some face identities in
the multi-party model averaging experiments (Section~\ref{sec:multi-avg}).
If for whatever reason the model does not internally separate the features
associated with the target property, inference will fail.

\subsection{Attribution of inferred properties}
\label{sec:attribution}

In the two-party scenarios considered in Section~\ref{sec:experiments2},
attribution of the inferred properties is trivial because there is only
one honest participant.  In the multi-party scenarios considered in
Section~\ref{sec:experimentsM}, model updates are aggregated.  Therefore,
even if the adversary successfully infers the presence of inputs with a
certain property in the training data, he may not be able to attribute
these inputs to a specific participant.  Furthermore, he may not be
able to tell if all inputs with the property belong to one participant
or are distributed across multiple participants.

In general, attribution requires auxiliary information specific to the
leakage.  For example, consider face identification.  In some applications
of collaborative learning, the identities of all participants are known
because they need to communicate with each other.  If collaborative
learning leaks that a particular person appears in the training images,
auxiliary information about the participants (e.g., their social networks)
can reveal which of them knows the person in question.  Similarly,
if collaborative learning leaks the authorship of the training texts,
auxiliary information can help infer which participant is likely to
train on texts written by this author.

Another example of attribution based on auxiliary information is described
in Section~\ref{sec:emerge}.  If photos of a certain person first appear
in the training data after a new participant has joined collaborative
training, the adversary may attribute these photos to the new participant.

Note that leakage of medical conditions, locations, images of individuals,
or texts written by known authors is a privacy breach even if it
cannot be traced to a specific participant or multiple participants.
Leaking that a certain person appears in the photos or just the number
of people in the photos reveals intimate relationships between people.
Locations can reveal people's addresses, religion, sexual orientation,
and relationships with other people.

\section{Related Work}\label{sec:related}

\paragraphbe{Privacy-preserving distributed learning.} 
Transfer learning in combination with differentially private (DP)
techniques tailored for deep learning~\cite{abadi2016deep} has
been used in~\cite{papernot2016semi,papernot2018scalable}.  These
techniques privately train a ``student'' model by transferring,
through noisy aggregation, the knowledge of an ensemble of
``teachers'' trained on the disjoint subsets of training data.
These are centralized, record-level DP mechanisms with a trusted
aggregator and do not apply to federated or collaborative learning.
In particular,~\cite{papernot2016semi,papernot2018scalable} assume
that the adversary cannot see the individual models, only the final
model trained by the trusted aggregator.  Moreover, record-level DP
by definition does not prevent property inference.  Finally, their
effectiveness has been demonstrated only on a few specific tasks (MNIST,
SVHN, OCR), which are substantially different from the tasks considered
in this paper.

Shokri and Shmatikov~\cite{shokri2015privacy} propose making gradient
updates differentially private to protect the training data.  Their
approach requires extremely large values of the $\varepsilon$ parameter
(and consequently little privacy protection) to produce an accurate joint
model.  More recently, participant-level differentially private federated
learning methods~\cite{mcmahan2017learning, geyer2017differentially}
showed how to protect participants' training data by adding Gaussian
noise to local updates.  As discussed in~\Cref{ssec:dp_defense}, these
approaches require a large number of users (on the order of thousands)
for the training to converge and achieve an acceptable trade-off
between privacy and model performance.  Furthermore, the results
in~\cite{mcmahan2017learning} are reported for a specific language model
and use \emph{AccuracyTop1} as the proxy, not the actual accuracy of
the non-private model.

Pathak et al.~\cite{pathak2010multiparty} present a differentially private
global classifier hosted by a trusted third-party and based on locally
trained classifiers held by separate, mutually distrusting parties.
Hamm et al.~\cite{hamm2016learning} use knowledge transfer to combine a
collection of models trained on individual devices into a single model,
with differential privacy guarantees.

Secure multi-party computation (MPC) has also been used to build
privacy-preserving neural networks in a distributed fashion.  For example,
SecureML~\cite{mohassel2017secureml} starts with the data owners (clients)
distributing their private training inputs among two non-colluding
servers during the setup phase; the two servers then use MPC to train
a global model on the clients' encrypted joint data.  Bonawitz et
al.~\cite{bonawitz2017practical} use secure multi-party aggregation
techniques, tailored for federated learning, to let participants encrypt
their updates so that the central parameter server only recovers the
sum of the updates.  In Section~\ref{sec:multi-avg}, we showed that
inference attacks can be successful even if the adversary only observes
aggregated updates.

\paragraphbe{Membership inference.} 
Prior work demonstrated the feasibility of 
membership inference from aggregate statistics, e.g., in the context
of genomic studies~\cite{homer2008resolving, backes2016membership},
location time-series~\cite{pyrgelis2017knock}, or noisy statistics in
general~\cite{dwork2015robust}.

Membership inference against black-box ML models has also been studied
extensively in recent work.
Shokri et al.~\cite{shokri2017membership} demonstrate
membership inference against black-box supervised models, exploiting
the differences in the models' outputs on training and non-training
inputs.  Hayes et al.~\cite{hayes2017logan} focus on generative
models in machine-learning-as-a-service applications and train
GANs~\cite{goodfellow2014generative} to detect overfitting and recognize
training inputs.  Long et al.~\cite{long2018understanding} and Yeom et
al.~\cite{yeom2017unintended} study the relationship between overfitting
and information leakage.

Truex et al.~\cite{demyst2018} extend~\cite{shokri2017membership}
to a more general setting and show how membership inference attacks
are data-driven and largely transferable.  They also show that an
adversary who participates in collaborative learning, with access to
individual model updates from all honest participants, can boost the
performance of membership inference vs.\ a centralized model.  Nasr et
al.~\cite{nasr2018machine} design a privacy mechanism to adversarially
train \emph{centralized} machine learning models with provable protections
against membership inference.

\paragraphbe{Other attacks on machine learning models.}
Several techniques infer class features and/or
construct class representatives if the adversary has
black-box~\cite{fredrikson2014privacy, fredrikson2015model} or
white-box~\cite{ateniese2015hacking} access to a classifier model.
As discussed in detail in Section~\ref{sec:privML}, these techniques
infer features that characterize an entire class and not specifically
the training data, except in the cases of pathological overfitting where
the training sample constitutes the entire membership of the class.

Hitaj et al.~\cite{hitaj2017deep} show that a participant in collaborative
deep learning can use GANs to construct class representatives.
Their technique was evaluated only on models where all members of
the same class are visually similar (handwritten digits and faces).
As discussed in Section~\ref{badprior}, there is no evidence that it
produces actual training images or can distinguish a training image and
another image from the same class.

The informal property violated by the attacks
of~\cite{fredrikson2014privacy, fredrikson2015model, ateniese2015hacking,
hitaj2017deep} is: ``a classifier should prevent users from generating
an input that belongs to a particular class or even learning what such an
input looks like.''  It is not clear to us why this property is desirable,
or whether it is even achievable.

Aono et al.~\cite{aono2017privacy} show that, in the collaborative deep
learning protocol of~\cite{shokri2015privacy}, an honest-but-curious
server can partially recover participants' training inputs from their
gradient updates under the (greatly simplified) assumption that the batch
consists of a single input.  Furthermore, the technique is evaluated
only on MNIST where all class members are visually similar.  It is not
clear if it can distinguish a training image and another image from the
same MNIST class.

Song et al.~\cite{song2017machine} engineer an ML model that memorizes
the training data, which can then be extracted with black-box access
to the model.  Carlini et al.~\cite{carlini2018secret} show that deep
learning-based generative sequence models trained on text data can
unintentionally memorize training inputs, which can then be extracted
with black-box access.  They do this for sequences of digits
artificially introduced into the text, which are not affected by the
relative word frequencies in the language model.

Training data that is explicitly incorporated or otherwise
memorized in the model can also be leaked by model stealing
attacks~\cite{tramer2016stealing,wang2018stealing,joon2018towards}.

Concurrently with this work, Ganju et al.~\cite{ganju2018property}
developed property inference attacks against fully connected,
relatively shallow neural networks.  They focus on the post-training,
white-box release of models trained on sensitive data, as opposed to
collaborative training.  In contrast to our attacks, the properties
inferred in~\cite{ganju2018property} may be correlated with the
main task.  Evaluation is limited to simple datasets and tasks such as
MNIST, U.S. Census tabular data, and hardware performance counters with
short features.

\section{Conclusion}
\label{sec:conclusion}

In this paper, we proposed and evaluated several inference attacks against
collaborative learning.  These attacks enable a malicious participant
to infer not only \emph{membership}, i.e., the presence of exact data
points in other participants' training data, but also \emph{properties}
that characterize subsets of the training data and are independent of
the properties that the joint model aims to capture.

Deep learning models appear to internally recognize many features of the
data that are uncorrelated with the tasks they are being trained for.
Consequently, model updates during collaborative learning leak information
about these ``unintended'' features to adversarial participants.
Active attacks are potentially very powerful in this setting because
they enable the adversary to trick the joint model into learning features
of the adversary's choosing without a significant impact on the model's
performance on its main task.

Our results suggest that leakage of unintended features exposes
collaborative learning to powerful inference attacks.  We also showed
that defenses such as selective gradient sharing, reducing dimensionality,
and dropout are not effective.  This should motivate future work on better
defenses.  For instance, techniques that learn only the features relevant
to a given task~\cite{edwards2015censoring,osia2017hybrid,osia2018deep}
can potentially serve as the basis for ``least-privilege'' collaboratively
trained models.  Further, it may be possible to detect active attacks that
manipulate the model into learning extra features.  Finally, it remains
an open question if participant-level differential privacy mechanisms can
produce accurate models when collaborative learning involves relatively
few participants.

\paragraphbe{Acknowledgments.}
This research was supported in part by the NSF grants 1611770 and
1704296, the generosity of Eric and Wendy Schmidt by recommendation of
the Schmidt Futures program, the Alan Turing Institute under the EPSRC
grant EP/N510129/1, and a grant by Nokia Bell Labs.\vspace*{-0.3cm}

\small
\bibliographystyle{abbrv}
\bibliography{citation}

\end{document}